\documentclass[twocolumn,amsmath,amssymb,prl,superscriptaddress]{revtex4}
\usepackage{graphicx}
\usepackage{bm}

\begin{document}
\title{Competing orders in FeAs layers}
\author{J.  Lorenzana} 
 \affiliation{SMC-INFM-CNR and Dipartimento di Fisica, Universit\`a di Roma
 ``La Sapienza'', Piazzale  Aldo Moro 2, 00185 Roma, Italy.}
\affiliation{ISC-CNR, Via dei Taurini  19, 00185 Roma, Italy.}
\author{G. Seibold}
 \affiliation{Institut f\"ur Physik, BTU Cottbus, PBox 101344,
          03013 Cottbus, Germany.}
\author{C.  Ortix}
\affiliation{Institute Lorentz for Theoretical Physics, Leiden University, P.O. Box 9506, 2300 RA Leiden, The Netherlands.}
 \author{M. Grilli}
 \affiliation{SMC-INFM-CNR and Dipartimento di Fisica, Universit\`a di Roma
 ``La Sapienza'', Piazzale  Aldo Moro 2, 00185 Roma, Italy.}
\date{\today}
\begin{abstract}
Using the unrestricted Hartree-Fock approximation and Landau theory we
identify possible phases competing with superconductivity in FeAs
layers. We find that close to half-filling the transition from the
paramagnet to the magnetically ordered phase is first-order making
anharmonicities relevant and leading to a rich phase diagram.  
Between the already known one dimensionally modulated magnetic stripe
phase and the paramagnet we find a new phase which has the same
structure factor as the former but in which magnetic moments
at nearest-neighbor sites are at right angles making electrons to 
acquire a non trivial phase when circulating a plaquette at strong
coupling. Another competing phase has magnetic and charge order and
may be stabilized by charged impurities.   
\end{abstract}
\maketitle
Exotic superconductivity often appears in a region where a
paramagnetic (PM) phase
competes with ordered phases. Examples are the heavy fermion
compounds\cite{mat98}, organics\cite{mck97,lef00,seo04}, vanadium
bronzes\cite{yam02} and barium bismuthates\cite{mat88,pei90}. 
A competing phase is believed to play an
important role in  Cu based high temperature 
superconductors\cite{zaa89,cas95b,kiv98,lor02b,sim02,wen96,cha01} probably
with charge\cite{tra95,abb05,koh07} and magnetic order\cite{tra95}
although there are other more exotic proposals\cite{sim02,wen96,cha01}. 
Alternatively one can take those as secondary effects and consider the
superconductor as emerging from the Mott
insulator\cite{kam90,zha97,cap02,and04} or from the Fermi liquid with 
antiferromagnetic correlations\cite{chu96}. 
One important question is if the competing phase is only
detrimental for superconductivity\cite{bal79}
or if the vicinity to an ordering instability helps 
superconductivity\cite{cas95b,kiv98,chu96,cap02}. 
 
In the recently discovered Fe-based superconductors, the Coulomb interaction is
believed to be relatively weaker than in cuprates making a mean-field
analysis more likely to lead to a quick identification of the possible
competing phases. Indeed first principle computation have predicted
a  magnetic stripe (MS) phase  with ordering wave vector 
$(\pi,0)$ (we define wave vectors for an isolated FeAs layer with 
the primitive vectors connecting nearest Fe sites and Fe-Fe distance 
$a\equiv1$) which later has been found by magnetic 
neutron scattering\cite{cru08}. 

Here we extend the analysis using the unrestricted Hartree-Fock
approximation in a model Hamiltonian which keeps only two orbitals
supplemented with a Landau theory analysis.  
This allows more flexibility than first principle computations to
identify possible competing phases at the expense of more (yet) unknown
parameters. Surprisingly, we find a
new magnetic phase in which magnetic moments at nearest neighbor sites 
are at right angles which we term orthomagnetic  (OM). This is
interesting in its own because well formed magnetic moments usually
lead to phases in which magnetic moments are either mutually parallel
or antiparallel\cite{xuc08}. We identify another competing  
phase which has both charge and spin order reminiscent to the stripe 
phases of cuprates. This phase may be locally stabilized by
charged impurities in the PM.

Our starting point is the two-orbital Hamiltonian $H=H_0 + H_{int}$
\cite{han08,li08,rag08,dag08}
with the non-interacting part (here written in momentum space)
\begin{eqnarray}
H_0&=&\sum_{k,\sigma}\left\lbrack\varepsilon_{k,1}d_{k,1,\sigma}^\dagger 
d_{k,1,\sigma} + \varepsilon_{k,2}d_{k,2,\sigma}^\dagger 
d_{k,2,\sigma} \nonumber \right. \\
 &+& \left. \varepsilon_{k,12}\left(d_{k,1,\sigma}^\dagger 
d_{k,2,\sigma} + d_{k,2,\sigma}^\dagger 
d_{k,1,\sigma}\right)\right\rbrack \\
\varepsilon_{k,1} &=& -2t_1\cos(k_x) - 2t_2\cos(k_y)-4t_3\cos(k_x)\cos(k_y) \nonumber \\
\varepsilon_{k,2} &=& -2t_2\cos(k_x) - 2t_1\cos(k_y)-4t_3\cos(k_x)\cos(k_y)  \nonumber\\
\varepsilon_{k,1} &=& -4t_4\sin(k_x)\sin(k_y) \nonumber
\end{eqnarray}
which describes the $d_{xz}-$ and $d_ {yz}-$ bands (hereafter ``$x$''
 and ``$y$'') on a square lattice
and the interaction part includes an intra- and
interorbital repulsion and a Hund coupling  term (here defined in real space):
\begin{equation}
H_{int}=U\sum_{i\alpha} n_{i\alpha\uparrow} n_{i\alpha\downarrow}
+ U'\sum_i n_{i x} n_{i y} \nonumber \\
- 2 J \sum_i{\bf S}_{i x}{\bf S}_{i y} .
\end{equation}
with $n_{i\alpha\sigma}$ the occupation
operator  for orbital $\alpha$ with spin $\sigma$,  $n_{i\alpha}\equiv n_{i\alpha\uparrow}+n_{i\alpha\downarrow}$ 
and ${\bf S}_{i \alpha}$ the spin operator for orbital $\alpha$.

In the following we adopt the hopping parameters from Ref.~\onlinecite{dag08}
and we define the intraorbital repulsion by the standard relation $U'=U-3J/2$
and $J=0.25 U$ for definiteness.

 Phases were identified using a fully  unrestricted 
HF approximation in large clusters (typically $14\times14$) with
periodic boundary conditions. Subsequently the energy of each solution
has been computed in much larger systems treated in momentum space. In
the latter case we neglected a small canting of the MS solution which
has a negligible effect on the energy.  

\begin{figure}[tbp]
\includegraphics[width=7 cm,clip=true]{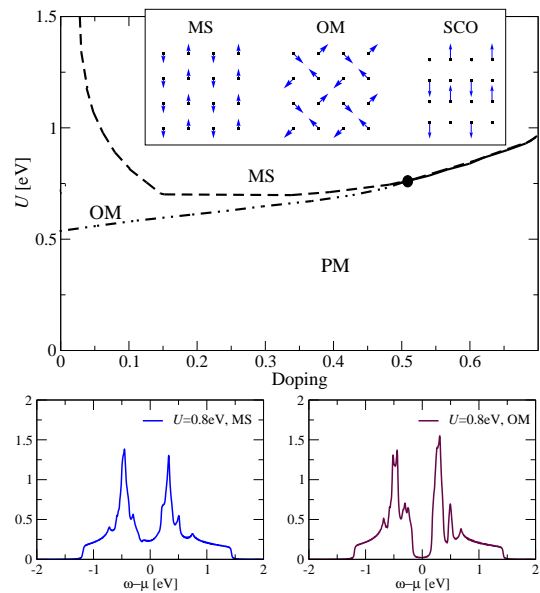}
\includegraphics[width=7 cm,clip=true]{dos1d2du08}
\vskip -0.3cm
\caption{(Color online) Upper panel: $T=0$ phase diagram of the
  two-orbital model in the $U$-density plane.  The solid circle is the
  tricritical point. The insets shows  the magnetic structure 
  of the ordered phases considered in this study. Lower Panels: The
  local HF DOS for the MS and OM states.}
\label{fig:phase2o}
\end{figure}

In the upper panel of Fig.~\ref{fig:phase2o} we show the resulting
phase diagram of the two-orbital model in the $U$-density plane at
zero temperature. The solid circle is a tricritical point: to the left
of it the transition is weakly first order whereas to the right it
is second order. Between the MS and the PM we find the new OM phase
in which magnetic moments at nearest-neighbor sites are at right
angles as shown in the inset of the upper panel of
Fig.~\ref{fig:phase2o}. It can be seen as a set of toroidal
moments\cite{ede07} mutually antiparallel thus it can also be termed an
``antiferrotoroidal'' state. 
It can also be viewed as the superposition of two magnetic stripes
perpendicular to each other. The magnetization at position ${\bf R}_l$
reads:
\begin{equation}
  \label{eq:mdl}
{\bf m}_l=\sum_{i=1}^2 {\bf M}_{i} \exp({\bf Q}_i.{\bf R}_l).   
\end{equation}
with ${\bf Q}_{1}\equiv ( 0,\pi)$,  ${\bf Q}_{2}\equiv(\pi,0)$ 
and ${\bf M}_{1}$,  ${\bf M}_{2}$ mutually perpendicular and equal in
modulus. The MS is recovered  by
setting  one of the  ${\bf M}_{i}$ in Eq.~\eqref{eq:mdl} 
to zero. 

The first order character of the transition close to half-filling is
not surprising given that the model is quite frustrated from the
magnetic point of view\cite{yil08}. Increasing $U$  the system can
avoid frustration by keeping in the PM phase until the point
in which the energy penalty of not forming  magnetic moments becomes too
large driving the system through the abrupt transition to a magnetic
phase.

We find the topology of the phase diagram shown in
Fig.~\ref{fig:phase2o} to be rather robust although for some
parameters the tricritical point becomes a triple point. Even more, we
find essentially the same topology in a standard one-band  Hubbard model 
supplemented with an explicit antiferromagnetic  next-nearest-neighbor 
magnetic interaction $J'$ across the diagonals. The latter  
frustrates the usual N\'eel state and favors the magnetic 
stability at ${\bf Q}_{i}$. 

The larger stability of the OM phase compared to the MS at half
filling  is due to the fact that this two dimensional texture opens
gaps in the electronic structure in both $x$ and $y$ direction whereas
the MS in weak coupling leaves the direction perpendicular to the
stripes metallic. Therefore a more stable fully gaped (insulating) 
solution is found for a weaker coupling in the OM case than in the MS case. 
This is illustrated in the lower panels of Fig.~\ref{fig:phase2o} 
where we show the density of states (DOS) for both solutions and
$U=0.8$eV. For smaller $U$ both solutions are metallic but the OM
state has deeper pseudogap.

The dips in the OM DOS close to 1/4 filling and 3/4 filling
can be understood in strong coupling as due to 
the fact that as an electron from the lower (upper) Hubbard band 
circulates a plaquette, the HF potential forces  its spin to be
parallel (antiparallel)  to the local direction 
of the magnetization. Thus after a complete loop the electron spin gets back to
the original direction but the single particle wave function acquires
a  phase $e^{i \pi}=-1$  reminiscent of the staggered flux phase in
cuprates\cite{wen96,aff88}. This produces a cone-like dispersion and a
semimetallic DOS which evolves into the dips in weak coupling.

More insight on the phase diagram can be obtained with the
following Landau toy model:
\begin{eqnarray}
\label{eq:gl}
\delta f&=&\frac12\sum_{{\bf q} } 
\chi^{-1}\left({\bf q}\right)  {\bf m}_{{\bf q}}.{\bf m}_{-{\bf q}}\\ 
&+&\beta_1 \sum_{{\bf q}_{1}...{\bf q}_{4}} 
({\bf m}_{{\bf q}_{1}}.{\bf m}_{{\bf q}_{2}}) ({\bf m}_{{\bf
    q}_{3}}.{\bf m}_{{\bf q}_{4}})  \delta_{ {{\bf q}_{1}+{\bf
      q}_{2}+{\bf q}_{3}+{\bf q}_{4}, 0}} \nonumber\\
&+&\gamma_1 \sum_{{\bf q}_{1}...{\bf q}_{6}} 
({\bf m}_{{\bf q}_{1}}.{\bf m}_{{\bf q}_{2}}) 
({\bf m}_{{\bf q}_{3}}.{\bf m}_{{\bf q}_{4}})  
({\bf m}_{{\bf q}_{5}}.{\bf m}_{{\bf q}_{6}})\nonumber\\
 &\times& \delta_{ {{\bf q}_{1}+{\bf q}_{2}+{\bf q}_{3}+{\bf
       q}_{4}+{\bf q}_{5}+{\bf q}_{6}, 0}} 
\nonumber 
\end{eqnarray}
It consists of a Gaussian part and a local
interaction part which we keep up to sixth order to be able
to describe both second- and first-order phase transitions. 
We assume that the susceptibility
 $\chi\left({\bf q}\right)$ diverges, for a set of symmetry related
 momenta ${\bf Q}_i$, at an instability line controlled by the parameter 
$\alpha\equiv\chi^{-1}\left({\bf Q}_i\right)$ and take $\gamma_1>0$
for stability. The non Gaussian vertexes have been taken as momentum
independent in the same spirit of Ref.~\cite{her76}. This will be
partially relaxed below. 

 For $\beta_1>0$ the above model describes an
ordinary second-order magnetic phase transition as $\alpha$ crosses the
Gaussian line. For $\beta_1<0$ there is a first-order phase transition
with a tricritical point at $\beta_1=0$. This can be treated
analogously to the liquid-solid transition\cite{ale78,cha95} and 
frustrated phase separation\cite{ort08}. As there, we restrict the sums
to the ${\bf Q}_i$ set that in the present case has only two
elements. Hence we have to choose subsets which satisfy the
Kronecker delta of the quartic term  modulo a reciprocal lattice
vector. Only two choices are possible:
either all the ${\bf q}_i$ are the same and equal to one ${\bf
  Q}_i$ which leads to the MS or the ${\bf q}_i$ are equal in 
pairs to one of the ${\bf  Q}_i$.  In the latter case we still have to
choose the relative angle between the two vectorial Fourier components 
${\bf M}_i$ at   ${\bf  Q}_i$  [c.f. Eq.~\eqref{eq:mdl}] and their
magnitudes, $M_i$\cite{rotational}.  
For two dimensional textures we find that the energy is locally minimized
when the magnitude of the two amplitudes is the same and the angle
between the two Fourier components is either $\pi/2$ (which leads to
the OM state) or zero. The latter is a new phase in which the real
space magnetization is zero in one sublattice  and forms an AF
structure in the other sublattice. This is the only
phase we have found in which the modulus of the magnetization is not
uniform. Since the charge density is a scalar, it couples with the
square of the magnetization. It follows that this phase will have both
spin and charge order (SCO) reminiscent of the charged stripe phases
in cuprates\cite{zaa89,lor02b}. In the present case the charge
ordering wave vector is $(\pi,\pi)$. It is suggestive that this
phase is similar to the charge ordering in BaBiO$_3$ 
which becomes superconducting when doped\cite{mat88,pei90}. 
In the following, for simplicity, we neglect the effect 
of the charge relaxation upon the energy which, in any case does not
affect the quadratic terms. Indeed we find in HF that this effect is small. 

Calling $M_T$ the magnitude of the magnetization in real space,
we find that the MS and the OM phase have the same energy\cite{energy} 
given by  $\delta f_{OM,MS}=\alpha M_T^2/2 + \beta_1 M_T^4 +\gamma_1
M_T^6$ while for the SCO phase  $\delta f_{SCO}=  \delta
f_{OM,MS}/2$. These
energies are easily rationalized from the fact that the non Gaussian
terms in Eq.~\eqref{eq:gl} are local in real space and in the SCO
state only half of the sites are magnetized 
(c.f. inset in Fig.~\ref{fig:phase2o}).

\begin{figure}[tbp]
\includegraphics[width=7 cm,clip=true]{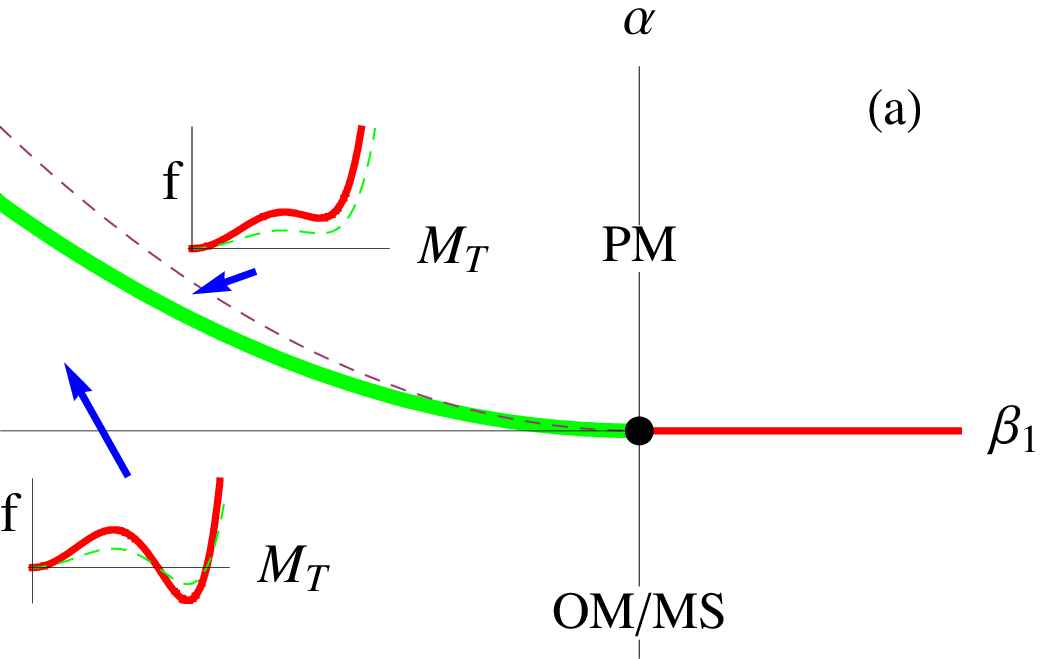}
\includegraphics[width=7 cm,clip=true]{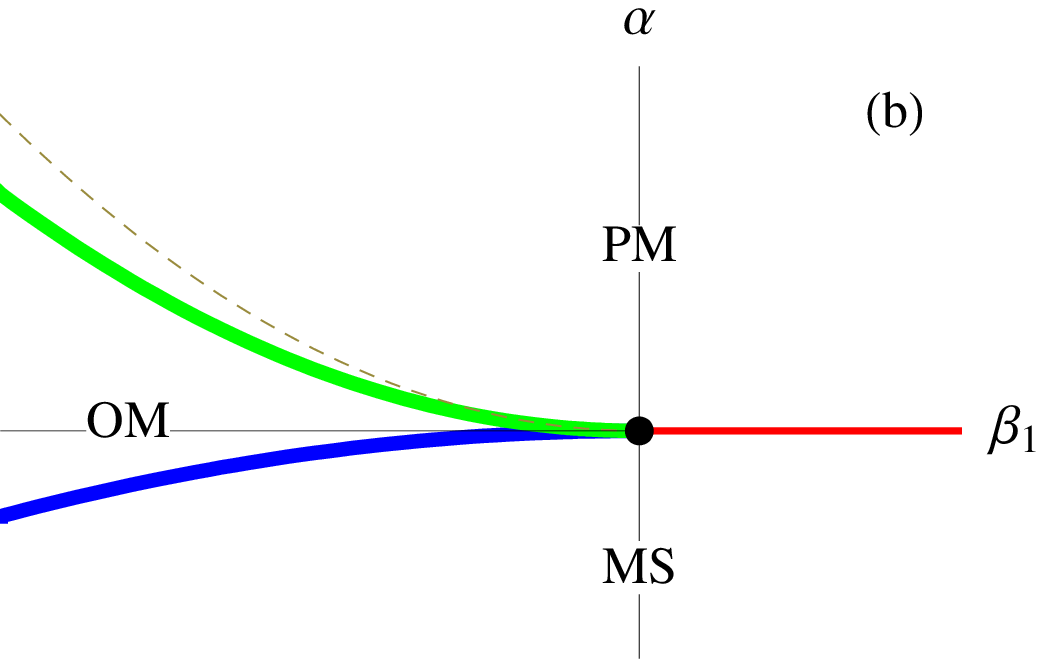}
\vskip -0.3cm
\caption{(Color online) 
 Phase diagram for the Landau toy model (upper panel) and the full
 Landau model (lower panel). 
The dashed line is the limit of metastability of
the ordered phases. The limit of metastability of the PM phase is
given by the line $\alpha=0$. The thick lines represent weakly first
 order transitions ending at the tricritical point. 
The insets in the upper panel  
show the behavior of the energy for the magnetic phases 
(full line) and the SCO phase (dashed line) in the regions of
 metastability. 
}
\label{fig:landau}
\end{figure}

The phase diagram is shown in  Fig.~\ref{fig:landau}(a).
The dashed line is the limit of metastability of the magnetic
phases. The insets show the behavior of $f$ in the metastable regions.  
The 1/2 factor in the energy makes the MS and OM configurations more
stable than the SCO configurations in the ordered region.  As soon as
the phase boundary is crossed and one reaches the PM phase 
the situation is reversed. The lowest energy phase above the PM is the
``hidden'' SCO phase.

In the restricted HF approximation for the parameters of
Ref.~\cite{dag08} and positive doping we have found that the PM-SCO
is second order, thus it can not become metastable as in the Landau toy
model but, of course this is sensitive to the parameter set. 
 For other parameters\cite{rag08} we have even found a region of SCO
phase in the phase diagram at high doping.

The Landau toy model phase diagram is of course oversimplified and does not
even reproduce the full physics of the HF phase diagram. One can
construct a canonical Landau theory as follows. 
First the magnetization is written as the sum of products
of slowly varying parts ${\bf M}_i({\bf R})$ times rapidly varying
parts  $\exp({\bf Q}_i.{\bf R}_l)$. The energy can be expanded
close to the Gaussian line in terms of the invariants generated by the 
${\bf M}_i({\bf R})$ and its gradients. Here we are interested in
uniform phases so we neglect the spacial dependence of the order
parameters ${\bf M}_i$ reaching Eq.~\eqref{eq:mdl}.

We have two second-order invariants $I_i\equiv {\bf M}_{i}.{\bf M}_{i}$. The 
fourth-order invariants are powers of the second-order ones and
$I_3\equiv ({\bf M}_{1}.{\bf M}_{2})^2$ and
$I_4\equiv ({\bf M}_{1}\times{\bf M}_{2}).({\bf M}_{1}\times{\bf M}_{2})$. The
latter can be eliminated since it depends on the other invariants
through the relation $I_3+I_4=I_1 I_2$.

The energy can be written as:
\begin{eqnarray}
\label{eq:landau}
\delta F&=&\frac12 \alpha \sum_{i=1}^2 {\bf M}_{i}.{\bf M}_{i}
+\beta_1 \sum_{i=1}^2 ({\bf M}_i.{\bf M}_i)^2+
\beta_2 ({\bf M}_1.{\bf M}_2)^2 \nonumber\\
&+&\beta_3 ({\bf M}_{1}.{\bf M}_{1})({\bf M}_{2}.{\bf M}_{2})\nonumber\\
&+&\gamma_1 \sum_{i=1}^2 ({\bf M}_i.{\bf M}_i)^3
+ \gamma_2 ({\bf M}_1.{\bf M}_2)^2  \sum_{i=1}^2 {\bf  M}_i.{\bf M}_i\\ 
&+&\gamma_3 \left[ ({\bf M}_1.{\bf M}_1) ({\bf M}_2.{\bf M}_2)^2+
  ({\bf M}_1.{\bf M}_1)^2 ({\bf M}_2.{\bf M}_2)\right]  
 \nonumber
\end{eqnarray}

In this case the energy of the different phases is:
\begin{eqnarray}
  \label{eq:ener_phases}
  \delta f_{MS}&=&\frac\alpha2 M_T^2 + B_1 M_T^4 + G_1 M_T^6\nonumber\\
  \delta f_{OM}&=&\frac\alpha2 M_T^2 + B_2 M_T^4 + G_2 M_T^6\nonumber\\
  \delta f_{SCO}&=&\frac12\left(\frac\alpha2 M_T^2 + B_3 M_T^4 +G_3 M_T^6\right)\nonumber
\end{eqnarray}
with $B_1=\beta_1$, $B_2=(2\beta_1+\beta_3)/4$, $B_3=(2\beta_1+\beta_2+\beta_3)/8$,
$G_1=\gamma_1$, $G_2=(\gamma_1+\gamma_3)/4$,
$G_3=(\gamma_1+\gamma_2+\gamma_3)/16$.

 The parameters of the model can be fixed by comparing
observables with experiment. One can consider the former to be function of two 
or more control parameters (like  $U$, $n$, $P$ and $T$). Taking two
control parameters for simplicity, each phase considered alone 
can have a different 
tricritical point with the paramagnet. If instead all phases are
allowed to compete, depending on parameters one can have more complex
phase diagrams including triple points. 
Given the local character of the electronic interactions, we expect the
parameters  to be not far from those of the Landau toy  model
Eq.~\eqref{eq:gl}, which corresponds to $B_i=\beta_1$, $G_i=\gamma_1$. 
Indeed one can obtain the same topology as in HF taking
$B_1=\beta_1$, $B_2=1.07\beta_1$, $B_3>\max(0,\beta_1/2)$, $G_2=1.1\gamma_1$ and
$G_1=G_3=\gamma_1$ [c.f. Fig.~\ref{fig:landau} (b)]. 

The four phases identified (PM, OM, MS, SCO) 
are the only phases allowed by the model Eq.~\eqref{eq:landau} thus
our study is exhaustive close to the Gaussian line (i.e. close to the
tricritical point in the first order region). The exception are  
special combinations of parameters where a family of
solutions become degenerate as in the 
 Landau toy model\cite{energy}.
Canting of the magnetic moments respect to the identified solutions
require higher order terms in the energy. 

It is easy to check that the structure factor averaged to 
take into account a superposition of randomly oriented domains
is the same for the OM and the MS phases, thus in principle it can be
difficult to distinguish among them with a magnetic neutron
scattering experiment alone on polycrystalline samples.
In practice, nuclear neutron scattering detects also a lattice 
distortion which breaks the $C_4$ symmetry of the lattice and 
probably stabilizes the MS\cite{yil08,cru08}. It is possible that 
the OM state can be stabilized suppressing the structural transition 
by pressure or chemical substitution.

The SCO phase may be locally stabilized by charged impurities
in the paramagnet. We expect  the concomitant spin and charge
order to appear around charged nonmagnetic impurities  which may be observed 
with local probes like NMR,  NQR and scanning tunneling spectroscopy.

In conclusion we have exhaustively analyzed possible magnetic phases
competing with  superconductivity  in FeAs layers close to the 
Gaussian instability line of the paramagnet.  
 Because of magnetic frustration we find a
tendency for PM-magnetic transitions to be first order as shown in the
HF approximation. This makes
non Gaussian terms relevant and generates an anharmonic order
parameter as shown in HF and using Landau theory. Contrary to 
general beliefs the two-orbital minimal model does not show the MS phase
close to half-filling. Instead we find
a new phase in which magnetic moments acquire an unusual orthogonal 
configuration. We find another low energy phase with spin order
and charge order at momentum $(\pi,\pi)$ which provides an obvious
link among charge fluctuations, possibly relevant for superconductivity,
and magnetism.  Our result provide a guide of likely configurations to
be found  in the phase diagram of layered FeAs  based compounds and a
Landau framework to study them.  
 

\begin{thebibliography}{10}
\bibitem{rotational}
Notice that because of spin rotational symmetry the energy is invariant with
  respect to a global rotation of the magnetizations.
\end{thebibliography}

\end{document}